# Tailoring superhydrophobic ZnO nanorods on Si pyramids with enhanced visible range antireflection property

D. Sharma,[1] S. Bhowmick,[2] A. Das,[3] A. Kanjilal,[2] and C. P. Saini[3,*]

[1]Department of Physics, Indian institute of Technology Delhi, Hauz-Khas, New Delhi 110016, India

[2]Department of Physics, School of Natural Sciences, Shiv Nadar University, NH-91, Tehsil Dadri, Gautam Buddha Nagar, Uttar Pradesh 201 314, India

[3] Inter-University Accelerator Centre, Aruna Asaf Ali Marg, New Delhi 110 067, India

## Abstract

Simultaneous superhydrophobic and visible range antireflective properties are demonstrated in hydrothermally grown ZnO nanorods (NRs) on chemically textured Si surfaces. A drastic transformation of Si micro-pyramids from hydrophobic to superhydrophobic is observed by securing the formation of polycrystalline ZnO NRs at surfaces, showing an increment of apparent contact angle from ~102° to ~157° and further explained in the light of a decrease in solid fractional surface area, according to the Cassie-Baxter model. Moreover, wurtzite phase of ZnO NRs is confirmed by the X-ray diffraction and Raman analysis. Such hierarchical structures further exhibit a large visible range antireflection (~5%), and correlated it to the variation in aspect ratio of Si pyramids in conjunction with the formation of graded refractive index. The combined superhydrophobic and large antireflection phenomenon in the present dual-scale structures is believed to be useful for Si-based photovoltaic applications.

[*]Author to whom correspondence should be addressed; E-mail: cs810@snu.edu.in

## 1. Introduction

Inspired form nature [1], artificially designed superhydrophobic surfaces (contact angle (CA) ≥ 150° and roll-of angle ≤ 10°) have attracted an immense interest in scientific community due to their potential to be used in wide range of applications such as self-cleaning [2], controllable oil/water separation [3], antibacterial activity [4], light harvesting [5] and so on. Beside chemical treatment, superhydrophobic property strongly depends on surface roughness, and this can be maximized by making porous or solid-air composite surfaces via development of micro- and nano-structures [6]. In particular, creating dual-scale hierarchical configuration with both micro/nano-structures (i.e. lotus leaf structure) shows high apparent CA [7]. The wettability of water on solid surfaces generally discuss in the light of simple Wenzel and/or Cassie-Baxter model [8], but they are unable to explain the phenomenon independently on complex structures. However, for developing artificial hierarchical structures, various approaches has recently been employed such as lithographic patterning [9], vapor-liquid-solid method [10], chemical route on self-assembled textures [11] as well as an cost effective chemical etching process [5, 12] and so on. Among them, chemical etching is more advantageous especially for Si which is a backbone of semiconductor industries because of their compatibility and cost effectiveness with respect to other competitors. In fact, the combination of nano- and micro-structures of Si can give better reliability by preventing the surface from being contaminated, together with repelling water under hostile environments[5, 13]. Such dual-scale structures also enhance the photovoltaic conversion efficiency by reducing reflection loss due to multiple scattering of incident light [5, 13]. Here development of subwavelength scale chemically textured Si pyramids allow the additional trapping/absorption of solar spectrum by diffuse scattering [14] and/or mode coupling [15]. However, in order to achieve dual-scale surface roughness with enhance self-cleaning property, Si pyramids are required to tune further by modifying the textured surfaces with suitable nanostructures such as nanorods (NRs) and/or nanowires etc. [5, 11] Among various parameters to integrate such nanostructures on micro-structured Si surfaces [12, 16], the proper material selection along with cost effective fabrication process for developing nanostructures is the key to improve super-hydrophobicity and antireflection properties without compromising device performance. In this respect, chemically synthesized ZnO NRs with suitable height on textured Si surfaces not only offer the self-cleaning properties, but also act as an anti-reflective layer (ARC), a prerequisite for enhancing the performance of a solar absorber by reducing the reflection loss [17].

In this letter, we investigate the multifunctional properties of hierarchical structure consisting of hydrothermally grown ZnO NRs on submicron facets of Si pyramids. In particular, we have shown that chemically textured Si pyramids can suppress the average surface reflectance below ~15%, particularly in the wavelength range of 500–1000 nm, though it is reduced further (~5%) by incorporation of ZnO NRs. It has also been revealed that such dual-scale hierarchical surface enables excellent superhydrophobic property with apparent CA of ~157°, a suitable step for removing dust particles from solar cells.

## 2. Experimental

Ultrasonically cleaned pieces of 500 μm thick $p$-type Si(100) wafers (area 1×1 cm$^2$) were initially chemically textured for developing Si pyramids that act as templates for growing ZnO NRs. The texturing process was executed at 70 °C for 40 min in a 3 wt. % NaOH solution diluted with 10% isopropyl-alcohol (IPA); details can be found in Ref [17]. Further ZnO NRs have been synthesized on theses Si pyramids by using two step processes. Firstly about ~10 nm thick ZnO seed layer was deposited at room temperature by Radio frequency (RF) magnetron sputtering with a 50 W power supply. The deposition pressure was maintained ~5 mtorr whereas substrate-to-target distance was kept at ~10 cm. Highly pure (99.999%) oxygen and Ar gas were injected into the chamber with a flow rate of 12 sccm and 30 sccm, respectively.

Following this, ZnO NRs were fabricated on seed layer coated textured Si surfaces by hydrothermal process where seeded wafers were placed into autoclave containing the aqueous solution of deionized (DI) water (75 ml), zinc nitrate (25 mM) and haxamethylene tetramine (25 mM). The autoclave was than kept in furnace at ~110 °C for 5 hrs. Afterwards, samples were removed from the chemical solution, rinsed properly with DI water followed by ethanol. The as-grown ZnO NRs on textured Si surfaces were then annealed in air at ~325 °C for 45 min for improving crystallinity.

The phase formation and crystalline structure of ZnO NRs on Si(111) facets were identified by X-ray diffraction, XRD (Bruker, D8-Discover) using a Cu-$K_\alpha$ radiation ($\lambda = 0.154$ nm) followed by Raman spectroscopy (Rainshaw Invia), excited by using argon laser at 514 nm with 10 mW power. Moreover, the surface morphology was examined by scanning electron microscopy, SEM (Carl Zeiss) whereas reflectance spectra of the samples were recorded by using an integrating sphere attached in a UV–VIS–NIR spectrophotometer (Shimadzu Solid-Spec-3700) in the wavelength range of 300-1200 nm. The wetting behavior of Si pyramids before and after NRs formation was investigated by CA measurement system (Krüss GmbH, DSA-25). Here, data was recorded by using software-controlled pendant drop mode whereas the repeatability of recorded data across the surface was also confirmed on each sample.

## 3. Results and discussion

Typical cross-sectional SEM (XSEM) image of $Si_{Tex}$ are displayed in Fig. 1(a), showing the random distribution of Si micro-pyramids throughout the surface. Since the hydrophobicity and particularly anti-reflective properties depend strongly on the shape, distribution, and most importantly on the aspect ratio [ratio of average height (*d*) to the distance between two consecutive pyramids (*a*)] of such microstructures, the optimization of these parameters are the key to enhance the overall device performance [18]. Note that IPA plays a crucial role in the evolution of chemically etched pyramidal structure [19]. This acts as a surfactant for minimizing stickiness of hydrogen bubbles that are produced as a byproduct during chemical reaction owing to the increase in hydrophilic nature of newly etched Si surface. In fact, hydrogen bubbles behave like a virtual-mask between chemical solution and Si surface, and in turn control the reaction rate [19]. Close inspection of Fig. 1(a) further revealed that protrusions jutted out from the (111) facets of the sidewalls (indicated by downward arrow) in conjunction with clusters of small pyramids (marked by dashed rectangle). On the other hand, magnified view of one of pyramid as depicted in Fig. 1(b) also revealed the formation of very sharp apex with smooth and even facets. Previous theoretical works by Sai *et al.*[18] reported that the surface reflection loss can be effectively reduced by using submicron $Si_{Tex}$ surface with aspect ratio of *d*/*a* approximately around unity. Therefore, here, special attention was given to maintain the corresponding values of *d* and *a* of the order of ~2.5 and 2 μm by controlling the reaction time and temperature so that trapping of light can be maximized.[18] In fact, apart from some bigger structures, pyramids with height around 4-6 μm have generally been found in conventional *c*-Si solar cells [20], while in present case, both *d* and *a* are on the order of ~2.5–3 and 2–2.5 μm, respectively [see Figs. 1(a) and (b)].

Further, plan view SEM image of ZnO NRs on Si$_{Tex}$ is displayed in Fig. 1(c), showing the formation of highly dense NRs on top of Si(111) micropyramids. The morphology and distribution of such hierarchically grown NRs were found to be strongly depending on initial nucleation sites which can be controlled by optimizing the seed layer thickness and growth parameters like temperature, reaction time, concentration, etc. [21] High resolution SEM of rectangular dashed region in Fig. 1(c) further revealed that most of NRs were aligned vertically on the Si facets, indicating the anisotropic growth of NRs along the *c*-axis [Fig. 1(d)]. The average diameter and length [extracted form XSEM (not shown)] of these NRs were measured to be 35 ± 3 nm and 1.5 ± 0.3 μm, respectively.

The structure of ZnO NRs on chemically textured Si (Si$_{Tex}$) was systematically investigated by XRD using coupled $\theta - 2\theta$ mode in the $2\theta$ range of 20°–75° [see Fig. 2(a)], where $\theta$ is the Bragg's angle. As discerned from the spectra, the dominant peaks appearing at $2\theta$ ~31.7°, 34.4°, and 36.2° can be assigned to the reflections from the (100), (002), and (101) planes whereas relatively small intense peaks observed at $2\theta$ ~47.5°, 56.5°, and 62.8° are associated with the reflections from the (102), (110), and (103) planes of ZnO NRs, respectively [22]. All the ZnO peaks can be indexed to polycrystalline wurtzite structure (Zincite, JCPDS 5-0664).[22] Moreover, the dominant peak located at ~69.15° can be attributed to reflection from the (400) plane of underneath Si$_{Tex}$ substrate [23]. No peaks from impurity were detected in the recorded spectra, confirming the formation of high-quality ZnO NRs.

To understand the nature of crystalline quality of such ZnO NRs on Si$_{Tex}$, Raman measurement was carried out, where the recorded spectrum is shown in Fig. 2(b). Hexagonal wurtzite ZnO structure usually associate to the $C_{6v}^4$ space group and according to group theory, it has eight sets of optical phonon modes at Γ point, referred as four Raman active (A$_1$ + E$_1$ + 2E$_2$) modes, two Raman silent $2B_2$ modes and two infrared active ($A_1 + E_1$) modes. In addition, $A_1$ and $E_1$ modes further separated to two components [i.e. longitudinal optical (LO) and transversal optical (TO) modes] [24]. As evident from the spectrum, a prominent phonon peak at ~438 cm$^{-1}$ and relatively weak peak intensity at ~378 cm$^{-1}$ were detected, while these peaks can be attributed to the $E_2$ (high) and $A_1$ (TO) modes of wurtzite ZnO structure, respectively [25]. Moreover, no signature of $E_1$ (LO) mode at ~583 cm$^{-1}$ is observed in the spectrum, possibly associated with the formation of high crystal quality of ZnO NRs with negligible lattice defects like oxygen vacancy and Zn interstitial [25, 26]. Beside this, the Raman peaks at ~619 and ~669 cm$^{-1}$ have been assigned to be originated from underneath $Si_{Tex}$ surface and related to the combine effect of acoustic and optical phonons at different symmetry points of the Brillouin zone of Si lattice [27].

To comprehend the importance of hierarchical structures, especially the role of ZnO NRs on $Si_{Tex}$ in improving the overall superhydrophobicity, CA measurements were carried out systematically for 2 μl water droplet. For comparison, apparent CA was also measured on flat Si ($Si_{Flat}$), as depicted in Fig. 3(a), giving an average value of 41° ± 2° (right panel). It is worth mentioning here that the hydrogen-passivated Si surfaces are generally hydrophobic in nature, whereas the formation of native oxide at surfaces convert them to hydrophilic [28]. In contrast, the anisotropic chemical texturing was found to enhance the surface roughness via formation of highly dense Si micropyramids [see Fig. 3(b)] and in turn it increased the apparent CA up to 102° ± 2° (right panel), indicating the hydrophobic nature (150° ≥ CA ≥ 90°). In fact, such hydrophobic nature of bare Si pyramids can be explained by Wenzel's law [29], where the water droplet penetrates the grooves between the rough surfaces. This is expressed by $cos\,\theta^* = r\,cos\,\theta$, where $\theta^*$ is the measured CA on a rough surface, $\theta$ is the average CA on the corresponding flat surface, and *r* is the fractional surface area representing the ratio between the actual area and projected area. However, existence of large CA hysteresis of about ~25° suggests the presence of adhesive force between the surface of $Si_{Tex}$ and impregnated water droplet [30].

Interestingly, the combination of ZnO NRs with Si micropyramids, as illustrated in Fig. 3(c), leads to a drastic increase in apparent CA up to 157° ± 2° (right panel) along with lower CA hysteresis (≤ 10°). This is due to the development of a dual-scale hierarchical structures [21], and in turn transform the surface to superhydrophobic. The present radical increase in CA can be discussed in the framework of Cassie-Baxter law [29] and calculated by $cos\,\theta^* = f\,(1 + cos\,\theta) - 1$ where *f* is solid fraction, defined area fraction of the liquid-solid interface [6]. According to this relation, it is clear that reduction in the value of solid fraction *f* results the increase in observed apparent CA ($\theta^*$). Therefore, unlike bare-Si micropyramids, formation of hierarchical structures here lead to considerable reduction in the effective solid surface area and in turn influence the water droplet to sit only between solid-air composite asperities without wetting the surface [12].

Being good light scattering centers, focus is now to understand the role of ZnO NRs in tuning the total reflectance (*R*) when grown on $Si_{Tex}$. In order to understand the underlying mechanism, UV-VIS experiments have been carried out in a systematic way in the range of 300-1200 nm (Fig. 4). As discerned from the spectra, *R* on the order of ~40–50% for polished $Si_{Flat}$ surface (black curve) is reduced dramatically to ~15% after surface texturing (highlighted in red), indicating the enhancement in light trapping by multiplying the internal reflections. It is important to note that *R* plays a substantial role in photovoltaic devices as the internal quantum efficiency, IQE = EQE/(1-*R*), where EQE is the external quantum efficiency [31]. The observed reduction in *R* can be explained in the light of aspect ratio (*d/a*) of submicron pyramids.

According to the recent theoretical prediction,[18] $d/a$ value should be within the range of 1−1.5 in order to achieve a minimum $R$ with lower surface recombination velocity,[17] in good agreement with our $d/a$ ratio (1−1.3). As depicted in Fig. 4, such microscale pyramids although exhibit excellent visible range anti-reflective property, but it still shows a relatively high $R$ around 380-450 nm owing to the limitation of geometric optics effect [32]. However, decoration of ZnO NRs on Si$_{Tex}$ not only help to overcome this issue because of their strong absorption edge in UV region, but also support to form the graded reflective index. It is worth to note that $R$ of any material is strongly dependent on its refractive index ($n$) by following relation: [33]

$$n = \left(\frac{1+R}{1-R}\right) + \sqrt{\frac{4R}{(1-R)^2} - k^2},$$ where, $k$ ($k = \alpha\lambda/4\pi$) is extinction coefficient. Therefore, the refractive index of ZnO NRs [$n$ ($\lambda$=300-800 nm) ~2.2-1.8] here is laying between air ($n$~1) and $c$-Si ($n$~3.4), and thus leads to engineer a graded reflective index for suppressing the Fresnel reflection over a broad range of spectrum [25].

## 4. Conclusions

In conclusion, we report the efficacy of hydrothermally grown ZnO NRs on chemically textured Si facets towards the application for superhydrophobic and antireflecting coating. Unlike hydrophilic Si$_{Flat}$ surface, bare-Si micropyramids initially exhibit hydrophobicity with an apparent CA of ~102°, however the decoration of vertically aligned ZnO NRs on such faceted surfaces leads to a significant increment in CA upto ~157°. This increase in CA was found to be correlated with the reduction of solid fractional surface area due to formation of dual-scale hierarchical structures, and discussed in the framework of Cassie-Baxter model. Moreover, such ZnO NRs on Si$_{Tex}$ were also found to suppress the visible-range reflectance below ~5%, and explained in the light of size and aspect ratio of Si pyramids, in conjunction with the formation of graded refractive indices at the ZnO NRs/Si$_{Tex}$ interfaces. Such superhydrophobic hierarchical structures along with reduced visible range reflectance will be a benchmark for Si-based photovoltaic applications.

## 5. Acknowledgements

CPS would like to acknowledge the financial support received from the University Grants Commission (UGC). SB and AK would like to acknowledge the financial support received from Shiv Nader University and Alexander von Humboldt Foundation for purchasing the contact angle system. The help received from Dr. T. Eshwar (CEERI, Plani) is highly acknowledged.

**Figure captions**

**Figure 1.** (a) Cross-sectional SEM image of $Si_{Tex}$, where magnified view of one such pyramid is shown in (b). Whereas plane-view SEM image of ZnO NRs on $Si_{Tex}$ are shown in (c) and (d), respectively.

**Figure 2.** Typical XRD pattern (a) and complementary micro-Raman (b) spectrum of ZnO NRs on $Si_{Tex}$ after annealed at 325 °C for 5 hrs in air.

**Figure 3.** Schematic representation (left panel) and the corresponding result (right panel) of the apparent CA measurements for (a) $Si_{Flat}$ (b) $Si_{Tex}$ and (c) ZnO NRs decorated $Si_{Tex}$ (dual-scale hierarchical structure), respectively. Here, each scale bar represents 0.5 mm whereas the errors in the measured CAs are found to be within ± 2°.

**Figure 4.** Reflectance spectra of $Si_{Flat}$ (black), $Si_{Tex}$ (red) and annealed (~325 °C in air) ZnO NRs grown on $Si_{Tex}$ (blue).

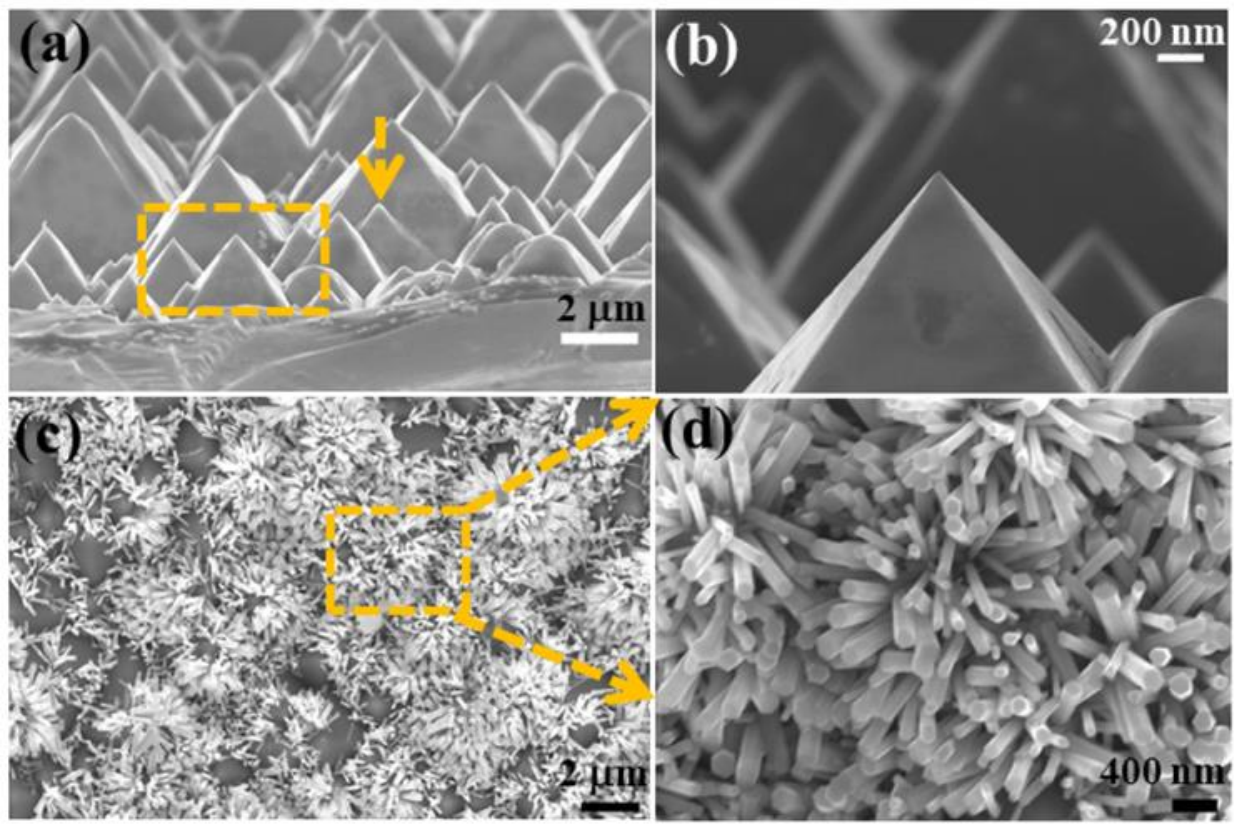

**Figure 1.** D. Sharma *et al.*

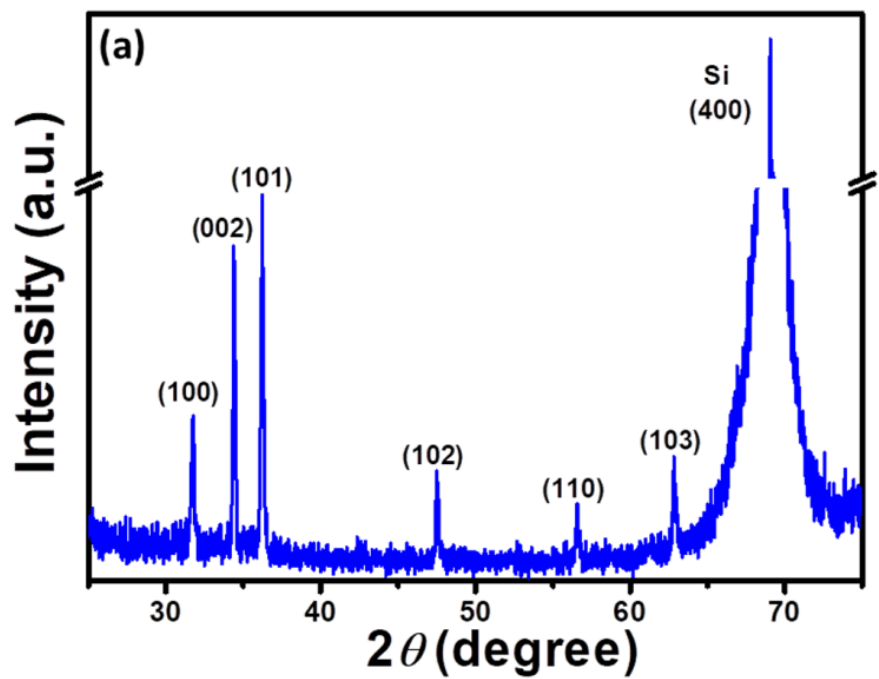

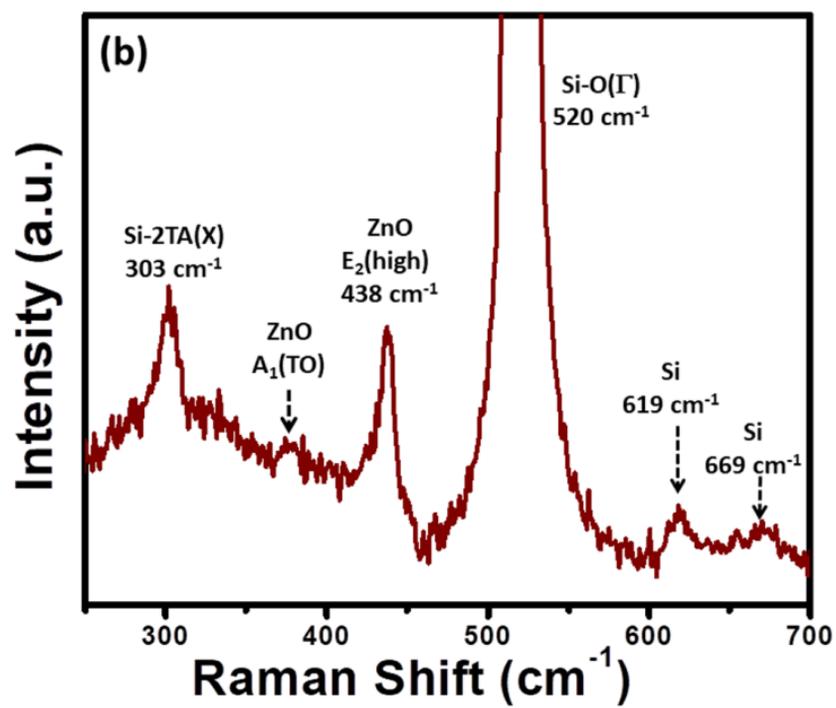

**Figure 2.** D. Sharma *et al.*

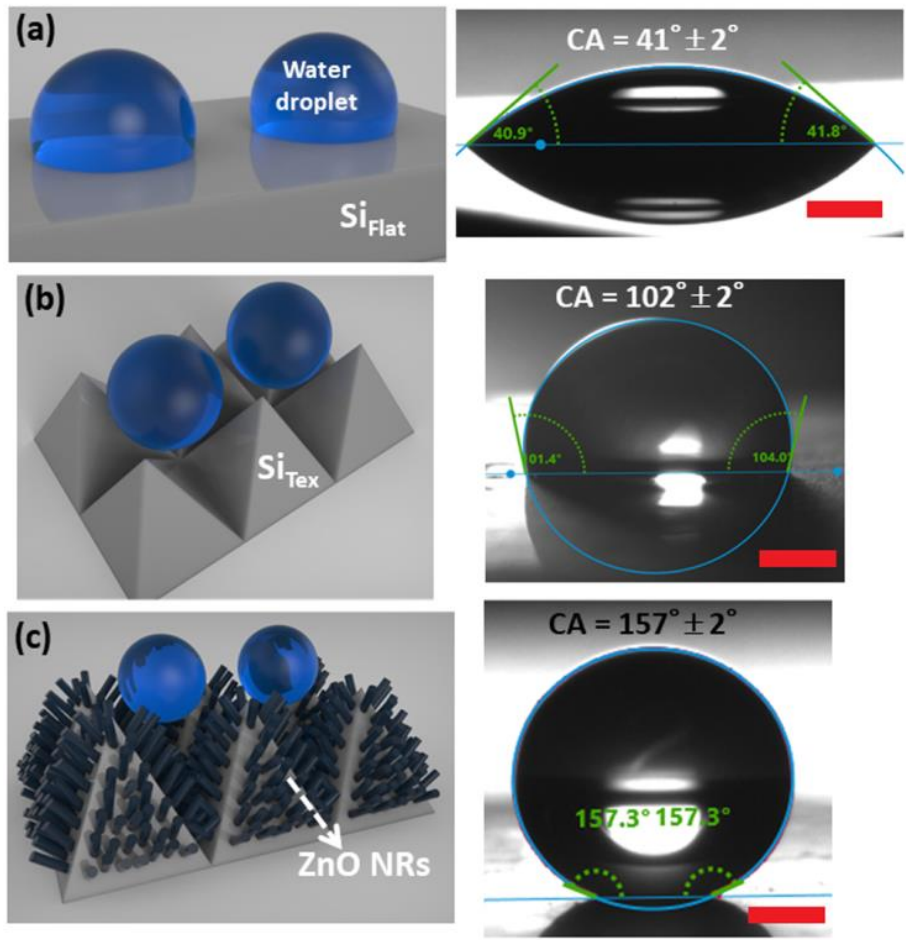

**Figure 3.** D. Sharma *et al.*

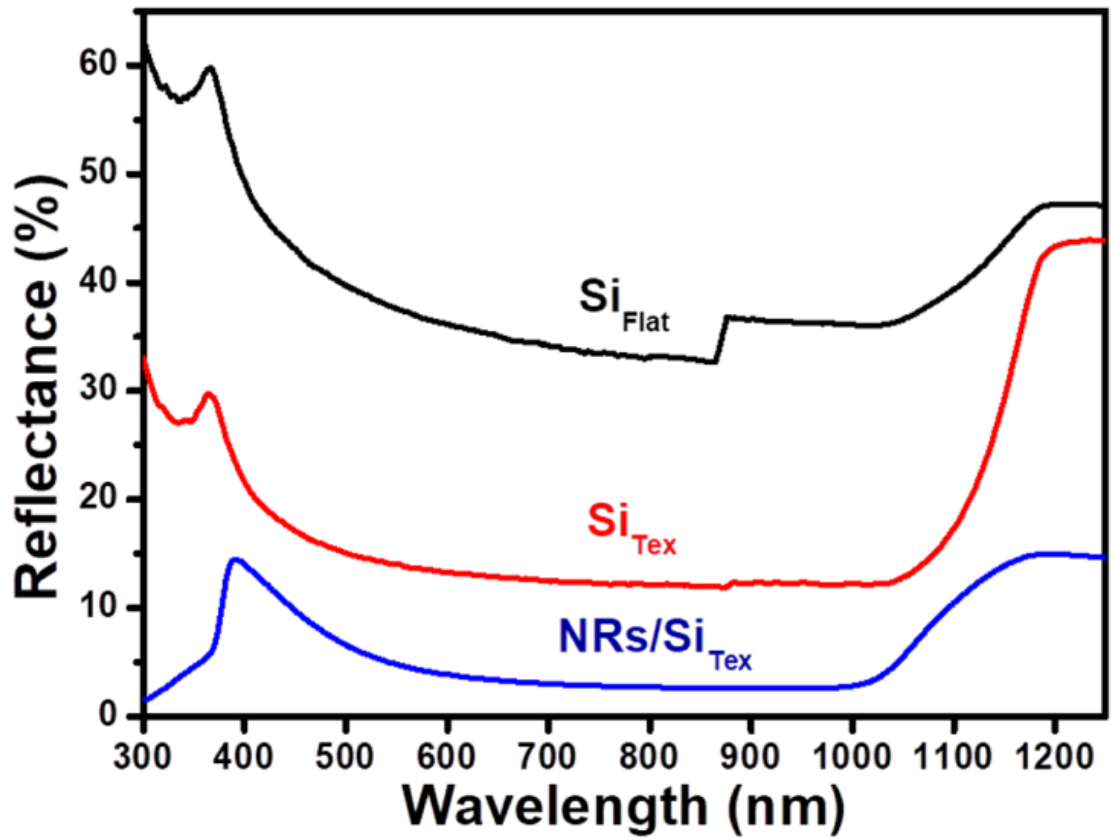

**Figure 4.** D. Sharma *et al.*